\def\Tr{{\rm Tr}}
\makeatletter
\renewcommand\@biblabel[1]{#1.}
\renewcommand\@cite[2]{$^{({#1\if@tempswa , #2\fi})}$}
\makeatother
\newfont{\blackboard}{msbm10 at 12pt}
\date{}

\def\cD{{\cal D}}
\def\cE{{\cal E}}
\def\cO{{\cal O}}
\def\cT{{\cal T}}
\def\cU{{\cal U}}
\def\cW{{\cal W}}
\def\twi{\~\,$\!$}
\def\lT{{\hbox{\blackboard T}}}
\documentclass[12pt]{article}

\usepackage{amssymb}
\title{Causality implies formal state collapse}
\author{George Svetlichny\thanks{Departamento de Matem\'atica,
Pontif\'{\i}cia Universidade Cat\'olica, Rua Marqu\^es de S\~ao Vicente 225,
22453-900 G\'avea, Rio de Janeiro, RJ, Brazil\hfill\newline {e-mail: svetlich@mat.puc-rio.br}\newline home-page: www.mat.puc-rio.br/\twi svetlich}}

\begin{document}
\def\skipu{12pt}
\parskip=\skipu
 
\maketitle
\begin{abstract} 
A physical theory of experiments carried out in a space-time region can accommodate a detector localized in another space-like separated region, in three, not necessarily exclusive, ways: 1) the detector formally collapses physical states across space-like separations, 2) the detector enables superluminal signals, and 3) the theory becomes logically inconsistent.  If such a theory admits autonomous evolving states, the space-like collapse must be instantaneous. Time-like separation does not allow such conclusions. We also prove some simple results on structural stability: within the set of all possible theories, under a weak empirical topology, the set of all theories with superluminal signals and the set of all theories with retrograde signals are both open and dense. 
\end{abstract}

\section{Introduction}

It has been said that quantum mechanics  exhibits ``spooky action-at-a-distance" through state ``collapse" across a space-like separation. We show that  collapse (whether considered spooky or not) is a formal feature of any complete, correct, and causal physical theory  which incorporates current experimental  practices. Thus said collapse is not a feature of quantum mechanics as such. This  offers an interesting twist on the usual discussions in the case of standard quantum theory. Space-like quantum state collapse, to the extent that it is not interpreted as a change of epistemic probabilities,  has always been considered a mysterious affair. It suggested to many people the existence of an action-at-a-distance and even that it could somehow be harnessed to carry superluminal signals. There have even been concrete proposals of superluminal communication devices based on this hope. From our perspective one can read the situation in a completely reverse fashion: that space-like quantum state-collapse has been empirically borne out by correlation experiments is not a suggestion that superluminal signals may be possible, but, being a necessary condition of their absence, is a suggestion of their impossibility. To our knowledge only Ryff and Monken\cite{Ryff} have come near this  point in showing that lack of collapse in certain ``interaction-free" measurements would lead to superluminal signals. The formal aspects of collapse in the case of quantum mechanics have also been  addressed by Macdonald\cite{AlMac}.

We argue our point by considering a theory that is already successful in describing phenomena in a certain space-time region and see how it must accommodate a detector acting in another region.  We have already used the artifice of a distant detector\cite{GS:linearity} to show that under some of the usual assumptions about quantum mechanical states, particularly the collapse hypothesis, part of the linear structure of quantum mechanics follows from the impossibility of superluminal signals. We see here to what extent we can relax the assumptions of the cited article and still draw significant conclusions.

Classical theories fall under our province and so have the same feature, but since up to now any collapse was always viewed incontestably as readjustment of the knowledge we have of the system, it was never seen as ``spooky" and generally not deemed a significant point to be discussed. 

Standard quantum theory, which should be used as a reference, is highly successful and highly accurate. If it is to be superseded by a new theory, this one must make numerical predictions that are very close to those for which the standard theory has been upheld in the laboratory, and so, in a certain sense, must be very close to quantum theory.  Our result thus does say something significant about our future views of  physics.

By a causal theory we mean one without superluminal communication and without retrograde signals (signals to the causal past). Specifically we disallow correlations in measurement results in different regions of space-time to result in the unwanted signals. We say that collapse is a formal feature to mean that the formalism must necessarily allow for such a description. If the actual discourse of physicists using the theory would contain such language, and if so, consider collapse a physical process, is a matter up to interpretation, taste, social practice, and metaphysics. Seeing as how twentieth century physicists have generally created their ontology directly from the formalism, such language would be probable, assuming no  radical shift of world view. 

Our discussion is mostly metatheoretic, that is, concerned with the mathematical and logical structure of physical theories. Mathematically the results are trivial, based on a reading of what formally is the conditional probability formula. Part of the argument is methodological as appeals are made as what is currently accepted experimental practice. 

Since the possibility of superluminal and retrograde signals is such a powerful metatheoretic device, it is desirable to know how prevalent these features are. Formalizing in a fairly primitive, but empirically significant fashion, the set of all possible theories, we point out that the subset of theories that have superluminal signals, and the subset of theories that have retrograde signals are both  dense and  open in the weak topology defined by a finite number of empirical observations.  If the absence of these features is part of Nature, any future theory transcending standard quantum mechanics would be exceedingly hard to construct. A striking exemplification of this last point are the results that show that absence of superluminal signals implies much of the linear structure of standard quantum mechanics\cite{GS:linearity, GS:cover,vry} provided we maintain some of the usual quantum mechanical notions about physical states. This fact is the main reason that no one has up to now presented a consistent relativistic non-linear quantum mechanics (See however Svetlichny\cite{nlrqm} and Kent\cite{kent} for initial attempts in this direction). Another consequence of the density of sets with these features is that any slight miscalculation in, or misunderstanding of, a causal theory could convince a person that she or he has stumbled upon a means of sending superluminal or retrograde signals. While quantum mechanics is still the prevailing theory we should continue to see more and more proposals to implement such signals using quantum mechanical collapse.

\section{Empirical Statistical Theories}\label{sec:et}

Physical theories, regardless of their conceptual and mathematical complexity or their metaphysical outlook on the world, must in the end describe and predict raw observational data. Different patterns in raw data would in principle require different theories to accommodate them, and a metatheoretic classification can be made on the basis of these patterns. Such a classification,  being to a large extent indifferent to the theories' mathematical and conceptual structures and to their metaphysical commitments, would have the greatest possible generality.

Any physical theory that proposes to predict the probabilities of experimental outcomes must at least admit preparation procedures that would allow one to perform  experiments repeatedly  ``under the same conditions'' and observation procedures that would allow
one to claim that one is ``performing the same experiment'' under these conditions. Any realizable experiment can lead to only a finite number of results and so if \(\cE\) is an observation procedure, we shall denote by \(\{\epsilon_1,\dots,\epsilon_n\}\)
its possible outcomes (``pointer positions''). Given a preparation procedure \(\cW\) and an observation procedure \(\cE\), the relative frequencies of outcome \(\epsilon_j\) in a long run is normally taken to approximate the {\em probability\/} \(P(\cW,\cE,\epsilon_j)\) of the result and which a scientific theory endeavors to calculate. Now both preparation and observation procedures are subject to some equivalence relations since it is generally accepted that different physical acts can
prepare the ``same state of affairs" and different apparatus can effect the ``same measurements'' on the said state of affairs.  Without going into any specific  analysis, such equivalence relations, which we denote by \(\sim\), must at least be statistically consistent, that is
\begin{eqnarray}\label{eq:weq}
 \cW \sim \cW' &\Rightarrow&\forall\,
\cE, \, P(\cW,\cE,\epsilon_j) = P(\cW',\cE,\epsilon_j) \\  \nonumber
\cE \sim \cE' &\Rightarrow& \exists \hbox{\ a permutation\ } \pi
\hbox{\ of\ }
\{1,\dots,n\} \hbox{\ such that \ } \\ \label{eq:eeq}
& & \forall\, \cW,\, P(\cW,\cE,\epsilon_j) = P(\cW,\cE',\epsilon'_{\pi(j)}).
\end{eqnarray}
Now the right-hand of the implications (\ref{eq:weq}) and (\ref{eq:eeq}) in themselves define equivalence relations that though implied by the one already postulated can be different. We call this possibly larger relation, {\em statistical
equivalence\/} and denote it by \(\approx\). The equivalence classes under \(\sim\) are generally part of what is formalized in a particular theory. 
In practice one usually says that the preparation procedure creates a ``physical
state'' which is subsequently ``measured'' by the observation procedure. Thus the equivalence classes of preparation procedures are often referred to as ``states'' and those of observation procedures as ``observables''. Such language is in accord with the
current physical intuition but is not at all necessary and we avoid it in general so as not to arouse ingrained semantic habits. 

 There must also be some sort of relativity theory that allows us to identify equivalent experiments performed in separate regions of space-time, and likewise some  sort of causality hypothesis to allow  {\em independently performed\/}  experiments. This is necessary to gather statistics on a ``run" of experiments.

We point out that while \(\approx\) is a purely empirical relation, \(\sim\) is a metaphysical concern. In quantum mechanics all three of the following prescriptions for creating a photon are \(\approx\) equivalent: preparation \(\cW_1\) is to pass light from an incandescent source through  a prism to pick out a given frequency;  preparation \(\cW_2\) is to pass light of the same frequency as in  preparation \(\cW_1\) (created in any way whatsoever) through a linear polarizer which is subject to random rotations from a horizontal to a vertical orientation;  preparation \(\cW_3\) is to create a two-photon singlet state twice the energy of  \(\cW_1\) and take one of the emerging photons as the prepared state. Now \(\cW_1\approx\cW_2\approx\cW_3\) and each, for computational purposes, is described by a \(2\times 2\) density matrix \(\frac12 I\). Now whereas \(\cW_1\sim\cW_2\) is likely to be maintained by most physicists, though some might contest it, there is some considerable disagreement about  \(\cW_1\sim\cW_3\) (and \(\cW_2\sim\cW_3\)). Some would say that \(\cW_3\) is not really a one-photon state  but a two-photon state in which one is merely doing a limited kind of measurement (only looking at one, which is ``unpolarized"). Others, however, feel that a beam of ``unpolarized photons" is ontologically the same as  a beam of photon of equally distributed random polarizations. 

Part of the notion of preparation and observation procedures is that the observation procedure is not to be considered as part of the preparation procedure. Thus the usual causal understanding is that the preparation is not influenced by any observation, for otherwise it would not be possible to talk about well defined states of affairs that one observes, nor of controlled experiments in general. 

Now one need not execute the observation procedure only in the instances that the preparation procedure is deemed successful. What does take  place in practice is that one performs a long experimental run and only {\em a posteriori\/} analyzes those instances in which the preparation was deemed successful. This is most apparent for instance in high-energy physics. A simple model for what happens in practice would be to consider that there is some ``gross" preparation procedure \(\cW_0\) and two observation procedures \(\cD\) with outcomes
\(\{\delta_1,\dots,\delta_p\}\), and  \(\cE\) with outcomes \(\{\epsilon_1,\dots,\epsilon_n\}\). A long experimental run is executed and only the cases in which \(\delta_1\) is realized are considered to be the cases in which the desired state of affairs has been created and for which the observations corresponding to \(\cE\) are then subsequently analyzed. Data for which some other outcome \(\delta_i\), \(i \neq 1\) obtains, is simply ignored. This is what is done for instance in coincidence experiments. The procedure described above can be called an {\em indirect preparation procedure\/}. The normal attitude concerning it is that the compound procedure: ``execute \(\cW_0\) then execute \(\cD\) and consider the
operation successful if \(\delta_1\) obtains", is just as legitimate a procedure for creating a state of affairs as any other. One still executes \(\cE\) and collects data even if \(\delta_1\) did not occur merely for technological reasons, it would just be too difficult or impossible to set up the experiment in another way. Since by assumption the separate executions of the experiment in the long run do not interfere with each other, the fact that the instances of the desired state of affairs are imbedded in a larger set along with states of affairs of no interest is innocuous as mere data analysis weeds them out. 

Observation procedures \(\cE\) and \(\cD\) may pertain to separate space-time regions which can be arbitrarily related to each other. If \(\cD\)  is time-like past to \(\cE\) then one has the usual {\em pre-conditioning} in which all of the preparation acts occur time-like prior to the observation acts. This is in accord with the usual causal notions. For space-like separation though, which acts are prior and which are subsequent, depends on the reference frame. This is accepted procedure and space-like conditioning is common in coincidence experiments. One  could even allow {\em post-conditioning\/}, that is conditioning a state to a future time-like measurement. Such practice, though perfectly legitimate, is not usually considered.  The bias against post-conditioning stems basically from conventional views concerning causality. Thus one does not generally admit that future time-like events condition present ones in some, admittedly ill-formulated, {\em physical\/} sense, and so while the mathematical and procedural status of post-conditioning is quite legitimate and uncontestable, one generally feels that for future time-like situations it has no physical basis and so any data analysis performed in this matter is totally superfluous and would not lead to any new physical insights. Summing up one can say that current experimental practices allow indirect preparation procedures, for both past-time-like pre-conditioning and space-like conditioning (either pre- or post-), but do not allow future-time-like post-conditioning.

Now the theory presumably has a certain mathematical formalization and predictive power based on a logical deductive system. In particular it should have formalized the \(\sim\) equivalence classes \([\cW]\) and \([\cE]\) of preparation and observation procedures and predict the values of the observed probabilities \(P(\cW,\cE,\epsilon_j)\) via  statements of the form 
\(P(\cW,\cE,\epsilon_j)=Q([\cW],[\cE],e_j)\).
Here \(e_j\) is the mathematical object that corresponds to the outcome \(\epsilon_j\). It also in principle predicts some ``laws of nature" which would be true predicates concerning the predicted probabilities \(Q([\cW],[\cE],e_j)\). We say a theory is {\em correct\/} if all predicted predicates (theorems) are in fact true on empirical facts and we say a theory is {\em complete\/} if every predicate true on empirical facts is in fact predicted (is  a theorem)  of the theory. 
If a theory is incomplete, expanding its axiomatic basis by the empirical facts that are not instances of theorems would turn it complete unless this procedure turned the theory incorrect (some new theorems are now in fact false). In this later case the theory certainly must be abandoned. We call a theory {\em successful\/} if it is correct and complete. Merely adding some facts to the axiom basis is phenomenology of the crudest sort, but is a necessary step to assure the overall consistency of the theory. We use the terms ``correct" and ``complete" in the purely logical sense of an interpreted deductive theory, not to be confused with other usage found in the physical literature.

Given a preparation procedure \( \cW\), then for any observation procedure \( \cE\) one has that \( \sum_iP (\cW,\cE,\epsilon_i )=1\). Let \(w\) be a map which to any \(\cE\) and to any  result \(\epsilon_i\) of \(\cE\) assigns a non-negative real numbers \(w^\cE(\epsilon_i)\). We say \(w\) is a {\em weight\/} if \(\sum_iw^\cE(\epsilon_i)\) is independent of \(\cE\), and we say the weight is {\em normalized\/} if this sum is one. Preparation procedures determine normalized weights through \(w^\cE_\cW(\epsilon_i) = P(\cW,\cE,\epsilon_i)\). We call such weights {\em physical\/}. Obviously a physical weight satisfies  \(\cE \sim \cE' \Rightarrow w^\cE_\cW(\epsilon_i) = w^{\cE'}_\cW(\pi(\epsilon_i))\), and  \(\cW \sim \cW' \Rightarrow w^\cE_\cW(\epsilon_i) = w^{\cE}_{\cW'}(\epsilon_i)\).
 
\section{Space-Time Extensions}\label{sec:sle}
 
Suppose now that we have a theory \(\cT\) in which the observation procedures all pertain to some bounded region of space-time \(\cO\). We assume this is a successful theory, that is,  it is correct and complete.  In contrast to the observation procedures,  do not assume that  the {\em preparation\/} procedures  pertain to \(\cO\), or for that matter to any specific space-time region. For instance, one may be observing some global cosmological structure.  Suppose now we have another space-time region \(\cO'\)  and that pertaining to this region is an observation procedure consisting of a detector \(\cD\) with  possible outcomes \(\{\delta_1,\dots,\delta_p\}\). 
We assume that we can run the detector in conjunction with any observation procedure \(\cE\) in \(\cO\). This, under normal causality assumptions, would always be the case if the two regions are space-like separated, but here we take it as a hypothesis and do not a-priori assume any space-time relations between \(\cO\) and \(\cO'\). We shall denote by \(\cO\nrightarrow\cO'\) the assumption that no signals can be sent from \(\cO\) to \(\cO'\) utilizing measurements in both regions and some preparation procedure. 

As far as the user of theory \(\cT\) is concerned, when he confronts the task of incorporating  detector \(\cD\), he is  faced with ``new phenomena" (``states" created by conditioning to outcomes of \(\cD\)),   ``new instruments" (concurrent measurements of \(\cE\) and \(\cD\)), and a ``greater world" \(\cO\cup \cO'\). 

The possible results of such a new compound experiment would be set of pairs \((\epsilon_i, \delta_j)\). We now ask what possible theories can there be of such compound procedures and what relation can these have to the original theory \(\cT\).

Let \(\hat\cT\) be any extension of \(\cT\) incorporating the detector \(\cD\) and let \(\hat P\) be the probability function for \(\hat\cT\), that is \(\hat P(\cW,\cE,\cD,\epsilon_i,\delta_j)\) is the joint probability of getting result \((\epsilon_i, \delta_j)\) using preparation procedure \(\cW\) and observation procedure \((\cE,\cD)\). Our first result is: 
\begin{equation}\label{eq:mtist}
\cO'\nrightarrow\cO\Rightarrow 
\sum_j\hat P(\cW,\cE,\cD,\epsilon_i,\delta_j) = P(\cW, \cE, \epsilon_i)
\end{equation} 
for otherwise one can, by choosing to run the detector \(\cD\) or not, send a signal from region \(\cO'\) to region \(\cO\). Thus \(\cO'\nrightarrow\cO\) implies that the {\em \(\cO\)-marginal\/} theory of \(\hat\cT\) in which one sums over the results of \(\cD\) has to coincide with \(\cT\). Now one in principle also has an \(\cO'\)-marginal theory \(\cT'\), though we have not postulated one. One can always formally consider it as a purely phenomenological one consisting of the same preparation procedures as \(\cT\), one observation procedure \(\cD\), the set of empirical facts given by  probabilities \(P'(\cW, \cD, \delta_j)\) of having outcome \(\delta_j\) in the {\em absence\/} of any observation in \(\cO\) and a trivial deductive system having no deduction rules. From the point of view of the user of \(\cT\) such a theory in itself is of no use as his task is to incorporate the new reality into \(\cT\) itself.  With this in mind we rewrite (\ref{eq:mtist}) for the reverse situation \(\cO\nrightarrow\cO'\) in a way that does not mention the \(\cO'\)-marginal theory.
This is our second result:
\begin{equation}\label{eq:cindx}
\cO\nrightarrow\cO' \Rightarrow \exists\, c_1\dots\exists\, c_p\,\forall\, \cE\, \sum_i\hat P(\cW,\cE,\cD,\epsilon_i,\delta_j) = c_j
\end{equation}
This holds since otherwise by changing to a different observation procedure \(\cE'\) one could send a  signal from \(\cO\) to \(\cO'\). Here \(c_j\) is the putative probability \(P'(\cW, \cD, \delta_j)\).

We note that if (\ref{eq:mtist}) holds, then   \(\cT\) can be maintained {\em under ignorance of the results of \(\cD\)\/}.  Regardless of (\ref{eq:mtist}) holding or not, 
theory  \(\hat\cT\) must meet various challenges under knowledge of the result of \(\cD\). Thus \(\sim\) may no longer hold statistically. It may happen  
that there are preparation procedures \(\cW \sim \cW'\) for which there is an \(\cE\) such that \(\hat P(\cW,\cD,\cE,\epsilon_i,\delta_j)\neq \hat P(\cW',\cE,\cD,\epsilon_i,\delta_j)\). Now \(\sim\) is  a metaphysical equivalence, so  a change of world view would allow one to pass to more fine-grained equivalence classes and recover statistical equivalence. Formally one may simply partition by the new \(\approx\),  each equivalence class \([\cW]\) in which the previous statistical equivalence is broken. It can also happen that 
there are observation procedures \(\cE \sim \cE'\) for which there is a \(\cW\) such that \(\hat P(\cW,\cE,\cD,\epsilon_i,\delta_j) \neq \hat P(\cW,\cE',\cD,\epsilon_{\pi(i)},\delta_j)\).
In general one can always perform the same trick of passing to finer-grained equivalence classes by partitioning any equivalence class \([\cE]\) which requires it, and adjusting our metaphysics to accompany this change. When  (\ref{eq:cindx}) holds, there is however another way to view the situation. Assuming   that \(c_j\neq 0\) then the map 
\begin{equation}\label{eq:wcond}
(\cE,\epsilon_i) \mapsto w^{\cE}_{(\cW,j)}(i) =
c_j^{-1}\hat P(\cW,\cE,\cD,\epsilon_i,\delta_j)
\end{equation}
is a normalized weight in \(\cT\) which respects the new \(\sim\) for  preparation procedures  and therefore has all the trappings of a physical weight. It is therefore either a physical weight, or one can incorporate ``conditioned to outcome \(\delta_j\)" as a legitimate new instruction in preparation procedures enlarging thus  the ``state space'' of \(\cT\).

This makes perfect sense for various reasons. The first is that \(c_j\) is independent of \(\cE\) and so one can indeed interpret indirect preparation procedure as a legitimate preparation for something to be measured since this very notion incorporates the idea of independence of preparation from the measurement. Secondly the number \(c_j\) is the putative probability that result \(\delta_j\) obtains in \(\cD\) so division by this number in (\ref{eq:wcond}) can be simply seen as normalizing the number of occurrences of \(\epsilon_i\) to the number of times that \(\delta_j\) has occurred as in usual conditional probabilities. Thus formally one has
\begin{equation}
w^{\cE}_{(\cW,j)}(i)=w^{\cE}_{\cW|j}(i)
\end{equation}
where \(\cW|j\) is the indirect preparation procedure alluded to. The  ``state space" is now enlarged to include the \(\cW|j\).  This can now be viewed as a ``conditioned state" and the theory can be said to have {\em formal state conditioning\/}, which we abbreviate it  SC.

As far as the deductive system  of \(\cT\) is concerned, under \(\cO\nrightarrow\cO'\) we can  enlarge the set of equivalence classes of preparation and observation procedures, and also enlarge the set of true empirical facts which we add to the axiomatic basis.  If the new resulting theory is successful we say we have {\em correctness stability\/} which we abbreviate to CS.

The  resulting successful theory now has 
formal state conditioning, and have thus proved that under \(\cO\nrightarrow\cO'\) one has
\begin{equation}\label{eq:slsc}
\hbox{CS\ }\Rightarrow\hbox{\ SC}
\end{equation}
If \(w_\cW\) is the physical weight corresponding to \(\cW\) then the map \(w_\cW \mapsto w_{(\cW,j)}\) where \(w_{(\cW,j)}\) is defined by (\ref{eq:wcond}) can be considered as indicative of a ``state collapse'' \([\cW]\mapsto [\cW|j]\) in analogy with the collapse hypothesis of standard quantum theory. Thus a form of state collapse is a consequence of \(\cO\nrightarrow\cO'\) and CS.  

Given the generality of our exposition, \([\cW|j]=[\cW] \) is a logical possibility. Though we may not want to consider ``no collapse" as ``collapse", we cannot in general
rule out that probabilities will not change under conditioning. Since any viable theory must necessarily be empirically close to quantum mechanics, change of probabilities must be as prevalent in it as in quantum mechanics, so ``collapse" as a generic description is justifiable.

\section{Space-Like Separation}

Here we assume \(\cO\) and \(\cO'\) are space like separated. Under usual causality assumptions \(\cO\nrightarrow\cO'\) and \(\cO'\nrightarrow\cO\) both hold. This precludes superluminal signals (SS) by use of the measurement process.
 Under \(\neg\hbox{SS}\) we have the conclusions of both (\ref{eq:mtist}) and (\ref{eq:cindx}). Thus the space-like variant of (\ref{eq:slsc}) is:
\begin{equation}\label{eq:slscp}
\neg\hbox{SS} \hbox{\ and CS\ }\Rightarrow\hbox{\ SC}.
\end{equation}
Which can also be written as the disjunct:
\begin{equation}\label{eq:trilemma}
\hbox{SC}\quad\hbox{or}\quad\hbox{SS}\quad\hbox{or}\quad\neg\hbox{CS}.
\end{equation}
A rough paraphrase of this would be that theory \(\hat\cT\) either  admits formal space-like state conditioning or else there are superluminal signals or else it is wrong. Note that this is true for the space-like extension {\em of any theory whatsoever\/} and so is a complete classification.

Of course one does not in practice develop a theory for one space-time region \(\cO\) only to then try to extend it to include other regions. One generally proposes some universal theory \(\cU\) of which both \(\cT\) and \(\hat\cT\) are to be restrictions. This means that any such proposal can be examined in the light of (\ref{eq:trilemma}) which now refers to \(\cU\). Since presumably \(\cU\) is a successful theory we are  then  left with the disjunction \(\hbox{SS or SC}\). Paraphrasing:

{\em
Any successful physical theory either allows superluminal signals or is  formally a space-like state-collapse theory. In particular, a successful causal theory necessarily exhibits formal space-like state collapse\/}.

\section{Theories With Autonomous Evolving States}

We now reexamine the above results for theories that admit the notion of {\em autonomous evolving state\/}. Given a preparation procedure \(\cW\) one assumes that at some instant \(t_{\cW}\) the procedure has run its course and has created a physical entity
\(\sigma_{\cW}(t_{\cW})\) usually called an {\em instantaneous physical state\/} and which from then on behaves autonomously without reliance on the procedure that gave it birth. One has the notion of {\em evolution\/} in that for \(t > t_{\cW}\) one has the state \(\sigma_{\cW}(t)\) which can be prepared by the following prescription: execute \(\cW\) and  wait \(t-t_{\cW}\) units of time. We shall consider the evolution to be deterministic in that given \(\sigma_{\cW}(t_{\cW})\), then \(\sigma_{\cW}(t)\)  is uniquely determined. For such a conception to be applicable to an observation procedure \(\cE\), this too must have associated to it a time instant \(t_{\cE}\) at which the apparatus ``engages'' the state and then it is a postulate of such theories that the probabilities of outcomes depend only on the instantaneous physical state at instant \(t_{\cE}\). Obviously if \( \cW \sim \cW'\), then by the meaning of \( \sim\) one should have for \( t > \max ( t_{\cW},t_{\cW'} )\) that \( \sigma_{\cW} (t )=\sigma_{\cW'} ( t )\). The probability functional \(P(\cW,\cE,\epsilon_i)\) can now be expressed entirely in terms of the instantaneous state \(P(\sigma(t_\cE),\cE,\epsilon_i)\) and one often conventionally drops all reference to preparation procedures and only deals with evolving autonomous physical states \(\sigma(t)\). 

We must emphasize that in these type of theories the states are considered as {\em instantaneous\/}. That is, the state description \(\sigma(t)\)  for \(t\ge t_{\cW}\) describes the state {\em at that instant\/} and refers to its entire supposed spatial extent, which may be the full simultaneity hyperplane in the frame of the observer. This is manifestly a frame dependent description and of course raises the well know issues of compatibility with relativistic covariance. As with relativistic quantum mechanics, we must assume that divergence of state descriptions maintained by observers in different reference frames does not lead to divergence of predicted experimental probabilities. We shall assume this requirement is met in a successful theory.

Let us now examine a theory with instantaneous physical states \(\cT\) pertaining to a space-time region \(\cO\) and again as in section \ref{sec:sle} assume we have a detector in a space-like separated region \(\cO'\) and that the extended theory \(\hat\cT\) is also one with instantaneous physical states. To analyze what \(\neg\hbox{SS}\) says in this case consider then a situation in which the detector \(\cD\) engages the instantaneous state \(\sigma(t)\) at some instant and that the detector outcome is realized  at an instant \(t_\cD\) before the apparatus of the observation procedure \(\cE\) engages it at instant \(t_\cE\). By the premises of instantaneous state theories, conditioning to outcome \(\delta_j\) of \(\cD\) must now define a new instantaneous state which comes into existence at an instant \(t_{\delta_j}\) with \(t_\cD \leq t_{\delta_j} \leq t_\cE\).  To conclude anything further one must invoke more hypotheses. Under conventional causality hypotheses, the space-like separated apparatus used in \(\cE\) cannot have any influence on what the description at \(\cD\) is and since one can run an appropriate such apparatus in such a way that \(t_\cE\) is arbitrarily close to \(t_\cD\), one must conclude that \(t_{\delta_j} = t_\cD\). In other words, absence of superluminal signals implies that \(\sigma(t) \) undergoes an {\em instantaneous\/} ``collapse" at \(t_\cD\) to a new instantaneous state \(\sigma_j(t_\cD)\) depending on the outcome \(\delta_j\) of the detector. Summarizing:

{\em Any successful causal theory describing autonomous evolving states, formally exhibits {\em instantaneous} space-like state-collapse.}

As applied to quantum mechanics, instantaneous state-collapse is at times interpreted as ``preparation-at-a-distance", that is manipulation of objective reality at a space-like separated location. Our result is a formal consequence of causality (to some extent contrary to such an interpretation) and must not be taken as implying any sort of action-at-a-distance. 

\section{Time-Like Separation}

There are two cases of time-like separation to consider, when \(\cO'\) is time-like past to \(\cO\), and when it is time-like future. In both cases, assuming the actual physical possibility of carrying out experiments \(\cE\) and \(\cD\), one can carry over some of the arguments in section \ref{sec:sle}. The difference stems from the fact that while signals from the future to the past may be deemed pernicious those from the past to the future are not bothersome and are in fact welcome as pyramids, filing systems, and memories attest.

In the first case, if one precludes retrograde signals, one has (\ref{eq:cindx}) and we can carry over the arguments of \ref{sec:sle} that follow that equation to conclude formal state collapse in successful causal theories in which the conditioning measurement \(\cD\) is time-like past to the observation procedure \(\cE\). This of course is the usual familiar situation. In the second case 
we can no longer claim the conclusion of (\ref{eq:cindx}). So  the sum in (\ref{eq:cindx}) can now depend on  \(\cE\) making  it impossible to consider the resulting state of affairs as being the result of a conditioning of the initial one by \(\cD\), independently of what \(\cE\) measurement one makes on it, precluding a state collapse interpretation.  The state of affairs is somehow mutually co-conditioned by both \(\cD\) {\em and\/} \(\cE\) and in general by any other measurement performed ``in between", that is in a region \(\cO''\) which is time-like future to \(\cO\) and time-like past to \(\cO'\). It escapes the usual paradigm of an experiment consisting of preparation and observation. Without any compelling argument that it offers new physical insights it can be generally ignored.

This situation occurs in fact in standard quantum mechanics. Assume  one has initially prepared a Heisenberg quantum state described by a density matrix \(\rho_0\) which is then subject to two successive measurements corresponding to self adjoint operators \(A\) and \(B\) each with discrete spectrum and corresponding finite spectral decomposition \(A = \sum_i \lambda_i P_i\) and \(B=\sum_j \mu_j Q_j\). According to the usual projection rule the joint probability of seeing outcomes \((i,j)\) for the two measurements is \(P(i,j)=\Tr (Q_jP_i\rho_0P_iQ_j)=\Tr(P_i\rho_0P_iQ_j)\) from which we derive the conditional probabilities \(P(j|i)\) and \(P(i|j)\). The first one is the pre-conditioned probability which is \(\Tr (Q_j \rho_i)\) where \(\rho_i = P_i\rho_0 P_i / \Tr(P_i\rho_0)\) is a new density matrix. Thus the state of affairs created by pre-conditioning is again expressible by a
density matrix which depends only on \(\rho_0\) and the projector \(P_i\) and so in particular is independent of \(B\), i.e. ``observation" does not interfere with ``preparation". The second, post-conditioned, probability,
\(P(i|j) = \Tr (P_i\rho_0P_iQ_j) /\sum_k\Tr
(P_k\rho_0P_kQ_j)\) 
is quite different in general whenever \(A\) and \(B\) do not commute. One cannot write \(P(i|j)\) as \(\Tr (P_i \rho_j)\) for any density matrix \(\rho_j\) determined solely by \(\rho_0\) and the spectral projectors of \(B\), as \(P(i|j)\) in general also depends  on the other compatible projectors \(P_k\), \(k \neq i\) that are measured along with \(P_i\). 

Given the above, it is not clear whether there  is any advantage in introducing post conditioned ``states of affairs" among what we consider to be our physical states and incorporate them into our theory. However, the case for this in  quantum mechanics was argued by Aharonov and collaborators\cite{Aha} who seem to claim that such new ensembles teach us something fundamental.

\section{Structural Stability}

Because any future fundamental theory must, on the level of raw experimental facts, be  neighboring to standard quantum theory, one is naturally led to investigate the neighborhoods of standard quantum theory in the set of all theories, provided these notions can be appropriately formalized.  We make a preliminary step in this direction and find that the set of theories with superluminal communication and the set of theories with retrograde signals are both dense open sets. This means that in any neighborhood of standard quantum mechanics there are radically different theories, and any  modification of quantum mechanics, no matter how small, can, if extreme care is not taken, produce theories with superluminal and retrograde signals.

As before, consider a theory with some preparation procedures  \(\cW\) and some  observation procedures  \(\cE\) with outcomes \((\epsilon_1,\dots,\epsilon_n)\).  The physical content of the theory is contained in the sets of numbers \(P(\cW,\cE,\epsilon_j)\).  The space of all possible theories that describe a certain fixed set of procedures for preparation and observations is the set of all possible such assignments. Since within the lifetime of any civilization, only a finite number of experiments can be performed, and each empirical probability can only be ascertained to a limited degree of accuracy, a statement of the empirical closeness of two theories \(P\) and \(Q\) would be of the form
\begin{equation}
|P(\cW,\cE,\epsilon_j)-Q(\cW,\cE,\epsilon_j)|<\epsilon 
\end{equation} 
where \(\epsilon >0\) and \(\cW\) and \(\cE\) run over some {\em finite\/} sets \(W\) and \(E\) respectively. For fixed \(P\), \(W\), \(E\), and \(\epsilon \) the set of \(Q\) that satisfies the above inequality is a neighborhood  of \(P\) in a topology that we call {\em weak\/}. Consider now a theory that can describe events in regions  \(\cO\) and \(\cO'\), and consider observation procedures with outcome sets (our generic \(\epsilon\)'s) being sets of pairs \((\alpha_i,\delta_j);\,i=1,\dots,n,\,j=1,\dots,m\) where \(\alpha_i\) refers to outcomes observed in \(\cO\) and \(\delta_j\) to outcomes observed in \(\cO'\).
Now for  \(\cO\nrightarrow\cO'\) to hold, it is necessary that  \begin{equation}\label{eq:isct}
\sum_i P(\cW,\cE,(\alpha_i,\delta_j))=\sum_\ell P(\cW,\tilde\cE,(\beta_\ell,\tilde\delta_k))
\end{equation} 
for all \(\tilde\cE\) that  coincide in their procedure with \(\cE\)   in region \(\cO'\) and where \(\delta_j\) and \(\tilde\delta_k\) correspond the same physically determined outcome in region \(\cO'\) for the two procedures. Intuitively speaking, this condition is a set of equalities, and equalities are easily broken by arbitrarily small perturbations, whereas strict inequalities are maintained for sufficiently small perturbations. One would thus expect that the set where a strict inequality holds to be dense and open. In terms of the assignments above, the only a-priori constraint is that \(P(\cW,\cE,\epsilon_j)\ge 0\) and \(\sum_jP(\cW,\cE,\epsilon_j)=1\) which defines for each pair \((\cW,\cE)\) a simplex \(\Delta_{(\cW,\cE)}\). So the set of all theories is the product \(\lT=\prod_{(\cW,\cE)}\Delta_{(\cW,\cE)} \) and it's clear that the weak topology is just the product topology of the usual topologies in the simplices. An equality (\ref{eq:isct}) is held on a proper linear subvariety of the product \(\Delta_{(\cW,\cE)}\times \Delta_{(\cW,\tilde\cE)}\) and so in \(\lT\) this defines a dense open set. Since the union of dense open sets is dense and open, any set of such equalities defines a set of theories whose compliment is dense and open. 

{\em Under the empirical notion of closeness, the set of theories with superluminal signals and the set of theories with retrograde signals are both open and dense.}

Of course \(\lT\) above is a very primitive description of the set of all theories. It doesn't for instance take into account the \(\sim\) equivalence relation, but to do so one simply goes over to the product over equivalence classes \(\prod_{([\cW],[\cE])}\Delta_{(\cW,\cE)} \) and precisely the same considerations apply. The conclusion may fail to hold if one imposed some further a-priori relation on the probability assignments or considered stronger topologies but to motivate these would require completely new metatheoretic and philosophic considerations.

\section{Acknowledgment}

The author thanks Alan~Macdonald and the referees for helpful comments. This work was partially supported by 
the Conselho Nacional de Desenvolvimento Cient\'{\i}fico e Tecnol\'ogico (CNPq).

\end{document}